\documentclass[draftcls,peerreview,12pt,onecolumn]{IEEEtran}
\usepackage{graphicx,amsmath,amssymb,subfigure}

\def\snr{\mathsf{SNR}}
 
\title{Advanced Coordinated Beamforming for the\\ Downlink of Future LTE Cellular Networks}
\author{George~C.~Alexandropoulos,~\IEEEmembership{Senior~Member,~IEEE}, \\Paul~Ferrand,~\IEEEmembership{Member,~IEEE}, Jean-Marie~Gorce,~\IEEEmembership{Senior~Member,~IEEE} \\and Constantinos B. Papadias,~\IEEEmembership{Fellow,~IEEE}
\thanks{G. C. Alexandropoulos and P. Ferrand are with Mathematical and Algorithmic Sciences Lab, France Research Center, Huawei Technologies Co$.$ Ltd$.$, 92100 Boulogne-Billancourt, France (emails: \{george.alexandropoulos, paul.ferrand\}@huawei.com).}
\thanks{J.-M. Gorce is with CITIlab-Inria, F-69621, Villeurbanne, France (e-mail:jean-marie.gorce@insa-lyon.fr).}
\thanks{C. B. Papadias is with Athens Information Technology, 151 25 Marousi, Athens, Greece (e-mail: cpap@ait.gr).}
}

\begin{document}
\maketitle
\begin{abstract}
Modern cellular networks in traditional frequency bands are notoriously interference-limited especially in urban areas, where base stations are deployed in close proximity to one another. The latest releases of Long Term Evolution (LTE) incorporate features for coordinating downlink transmissions as an efficient means of managing interference. Recent field trial results and theoretical studies of the performance of joint transmission (JT) coordinated multi-point (CoMP) schemes revealed, however, that their gains are not as high as initially expected, despite the large coordination overhead. These schemes are known to be very sensitive to defects in synchronization or information exchange between coordinating bases stations as well as uncoordinated interference. In this article, we review recent advanced coordinated beamforming (CB) schemes as alternatives, requiring less overhead than JT CoMP while achieving good performance in realistic conditions. By stipulating that, in certain LTE scenarios of increasing interest, uncoordinated interference constitutes a major factor in the performance of CoMP techniques at large, we hereby assess the resilience of the state-of-the-art CB to uncoordinated interference. We also describe how these techniques can leverage the latest specifications of current cellular networks, and how they may perform when we consider standardized feedback and coordination. This allows us to identify some key roadblocks and research directions to address as LTE evolves towards the future of mobile communications.
\end{abstract}

\newpage
\section{Introduction}\label{sec:Introduction}
Current dense and future super dense mobile broadband networks are subject to various scenarios of simultaneous interfering communication links.
In cellular networks, interference from neighboring base stations (BSs) is still one of the most prominent performance degradation factors resulting in outages or performance losses at the cell edges as well as increasing the need for complex handovers. A classical approach to tackle interference is through medium access control and medium sharing techniques, which in turn severely compromise the performance of each individual user in the network due to explicit time sharing over the common resources. As we move towards denser networks with BSs and access points covering smaller areas to get antennas closer to the users, interference is becoming increasingly challenging \cite{Jungnickel2014}.

Interference management in cellular networks has been first and foremost implemented through smart reuse of network resources, mostly through the so-called Frequency Division Multiple Access (FDMA) techniques. Previous generations of cellular network standards employed orthogonal \emph{reuse-$n$} schemes, where neighboring cells do not interfere on each others' resources. A frequency band used by a cell is not allowed, in this paradigm, to be used by neighboring cells, thereby greatly lowering the inter-cell interference floor. While the previous generation of mobile communications, namely Universal Mobile Telecommunications System (UMTS), moved from the reuse-$n$ to a reuse-$1$ paradigm, today's Long-Term Evolution (LTE) specifications include a more fine-grained approach \cite{Hamza2013}.
In classically deployed networks with large homogeneous cells, a core observation was that interference is mainly an issue for mobile terminal (MTs) laying far from their respective BSs, i$.$e$.$, at the cell edges. According to this approach, LTE BSs separate frequency bands dynamically and ensure that those allocated to the cell edges are non-overlapping. Such fractional frequency reuse (FFR) schemes are a very efficient form of interference management as it requires relatively low coordination from the BSs' part. On the other hand, it may require more advanced power control in the downlink, and from the network point of view, BSs inefficiently use the time and frequency resources.

Capitalizing on the wide deployment of multiple antennas, especially at the BS side, and the advances in multi-antenna signal processing techniques, a new approach for interference management has made its way into mobile communication standards. Coordinated Multi-Point (CoMP) \cite{J:Lee_ComMag} is a broad umbrella name for coordination schemes that aim at realizing multi-user communications, i$.$e$.,$ sharing the medium among multiple network nodes over space on top of the possible sharing over time and frequency resources \cite{Gesbert2010}. Focusing on the downlink and considering joint transmission (JT) CoMP, in the theoretical limit of infinitely many distributed antennas, one could exactly pinpoint each MT and ensure that the signal intended for it adds up at its position, while creating no interference for the other MTs in the network. In this case, interference is not only removed, but is actually harnessed and exploited to increase the received signal power at each MT. However, for the practical implementation of JT CoMP schemes, sharing of channel state information (CSI) and data for the targeted MTs among the coordinated BSs as well as tight synchronization at the data level among them are necessary. These requirements are actually constituting the major downfall of JT CoMP in practical cellular networks, rendering hard to achieve its theoretical gains in practice. On top of that, it was shown in \cite{Irmer2011, J:Lozano_Limits} that, imperfect and/or outdated CSI and uncoordinated interference have a very large impact on the performance of conventional JT CoMP schemes. Practical radio-frequency (RF) components, such as oscillators with phase noise, were also shown to have a similar effect \cite{Jungnickel2014}. 

As an alternative to JT CoMP for the downlink of cellular networks, Coordinated Beamforming (CB) is based on shared knowledge of the spatial channels between the coordinated BSs and their intended MTs to separate the different data streams without exchanging MTs' data. As such, CB schemes come with less stringent synchronization and coordination requirements \cite{Irmer2011}, while retaining at least a large part of the JT CoMP performance. With CB, coordinated BSs only share CSI, and as long as the CSI is up to date, synchronization is unneeded and each BS in the coordination cluster may transmit independently. Recent releases of the LTE specifications by the 3\textsuperscript{rd} Generation Partnership Project (3GPP) have integrated the necessary elements to estimate the interfering channels on the MT part, with added reference signals and coordination of these signals among the coordinated BSs \cite{3GPPIRCspecs,3GPPPhysProc}. 3GPP also included advanced 3-dimensional (3D) beamforming capabilities and more complex antenna patterns in the latest standards as well as associated simulation tools. Although the standardization of CSI exchange between BSs is still left to the discretion of the vendors, the aforementioned improvements enable the practical implementation of CB schemes, on which we focus the present article. The theoretical design of CB schemes has been lately the subject of many research papers, of which representative examples are \cite{J:Iterative_Jafar, J:Luo_TSP_IterMMSE, Suh2011, J:Alexandg_Recon2013}. Among these schemes, some \cite{J:Iterative_Jafar, J:Alexandg_Recon2013} target at the so-called multiple-input multiple-output (MIMO) \textit{interference channel (IFC)}, where each multi-antenna BS belonging to the coordination cluster wishes to serve exactly one multi-antenna MT, while \cite{J:Luo_TSP_IterMMSE, Suh2011} are intended to the more general MIMO \textit{interference broadcast channel (IBC)}, where each coordinated multi-antenna BS may serve concurrently more than one multi-antenna MTs.

In this article, we present comparative performance evaluation results among the recent CB schemes, which constitute future candidates for implementation in practical cellular networks due to their offered theoretical performance gains coming with reduced coordination overhead, and their increased level of compatibility to the latest relevant standards' specifications \cite{3GPPIRCspecs, 3GPPComPV11}. To advocate on the adequacy of interference coordination, only at the beamforming level, as an enabling approach for boosting the performance of dense networks, we consider as example scenarios of interest small-cell network deployments, where high capacity and tightly synchronized on the signal level links among the BSs belonging in a coordination cluster are not feasible. In such scenarios, coordination may be fully dynamic as a result of a scheduling mechanism, and hence, carried out through dedicated wireless links. We focus on revealing the potential resilience of the CB schemes \cite{J:Iterative_Jafar, J:Luo_TSP_IterMMSE, Suh2011, J:Alexandg_Recon2013} to uncoordinated interference and investigating their performance with standardized feedback \cite{3GPPPhysProc}. The latter goal may also serve as an indicator of the impact of the quality or latency of CSI to the performance of the considered schemes. To achieve the former goal, we propose a parametric system model where the powers of Intra-Cluster Interference (ICI) and Out-of-Cluster Interference (OCI) are defined relatively to the power of the desired signal. The impact of OCI on both the clustered and centralized CB schemes designed for the IFC, and on the decentralized schemes that can be applied to the IBC is assessed. We then discuss how to adapt these schemes in current and future standards, and how practical feedback and quantization may impact their performance. Finally, we conclude with some specific research directions, that may be pursued to improve the performance and integration of CB schemes in future LTE networks and beyond.

\section{Modeling Interference in Cellular Networks}\label{sec:Cellular_Model}
\begin{figure}[!t]
\centering
\includegraphics[keepaspectratio,width=6.5in]{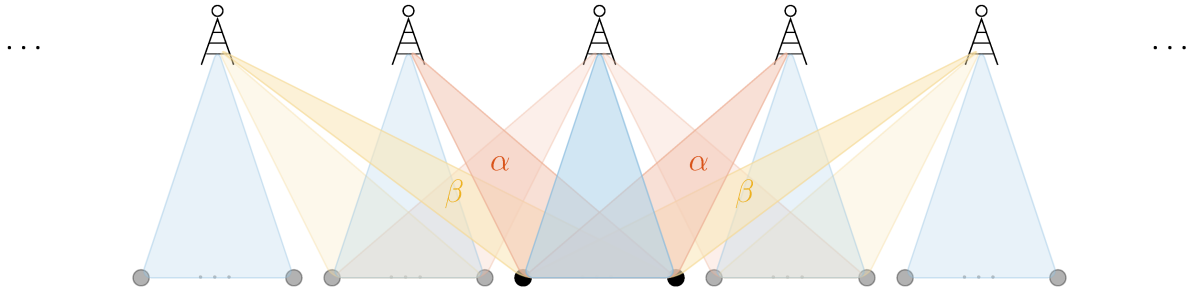}
\caption{A toy part of a cellular network comprising of $5$ BSs, each designed to service a group of MTs (on the bottom as black and grey dots). BSs are separated into two groups, with one of the groups including the $3$ BSs in the center that form a coordination cluster. Each MT associated to a BS belonging in the latter cluster is subject to ICI of relative power $\alpha$ (in red) as well as to OCI of relative power $\beta$ (in yellow).}
\label{Fig:General_IBC}
\end{figure}

To investigate the impact of interference in coordinated transmission schemes, we hereinafter present a simple system model that captures the relative effect of ICI and OCI in the received signal. For the interference experienced by each MT associated to a BS belonging to a coordination cluster, we make the following assumptions:  
\begin{itemize}
	\item The aggregate ICI is of relative power $\alpha\in[0,1]$ compared to the power of the desired signal.	For example, $\alpha=1$ models cases where MTs are at similar distances from all BSs in the coordination cluster. The latter case is well suited for MTs located at the edges of the separate cells, and thus the center of the cluster, where JT CoMP and CB schemes are expected to perform best.
	\item The aggregate OCI is of relative power $\beta\in[0,1]$ compared to that of the desired signal. This parameter indicates the effectiveness of BS clustering for coordinated transmission. Low values of $\beta$ indicate that most of the interference for a specific MT has been included within the cluster.	
\end{itemize}
Using the latter two assumptions, the proposed system model is mathematically described as follows. We consider an infinitely large cellular network from which we single out $B$ BSs, indexed in the set $\mathcal{B}=\{1,2\dots,B\}$, to form a coordination cluster.
On some time-frequency resource unit, the BS cluster aims at providing service to $U$ MTs indexed in the set $\mathcal{U}=\{1,2,\dots,U\}$. A set of MTs associated to BS $b\in\mathcal{B}$ is denoted by $\mathcal{U}_b$ such that, all sets $\mathcal U_b$ for all $b$ form a partition of the set $\mathcal{U}$. Without loss of generality, we assume that each BS is equipped with a $N$-element antenna array whereas, each MT has $M$ antennas. Let also $\mathbf{x}_{u,b}$ represent the $N$-dimension vector with the information bearing signal transmitted from the BS $b$ and intended for the MT $u$. Then, the baseband received $M$-dimension vector at the MT $u$ can be expressed as
\begin{equation}\label{Eq:Received_Signal}
\mathbf{y}_u = \mathbf{H}_{u,b}\mathbf{x}_{u,b} + \sum_{k\in\mathcal{U}_b,k\neq u}\mathbf{H}_{u,b}\mathbf{x}_{k,b} + \sum_{\ell\in \mathcal B, \ell\neq b} \sum_{n\in\mathcal{U}_{\ell}}\mathbf{H}_{u,\ell}\mathbf{x}_{n,\ell} + \mathbf{g}_u + \mathbf{n}_u
\end{equation}
where $\mathbf{H}_{u,b}$ denotes the $M\times N$ channel matrix between the MT $u$ and the BS $b$, and the $M$-dimension vector $\mathbf{g}_u$ is the OCI, for which we model the amplitude of its elements as independent and identically distributed Nakagami-$m$ random variables. It can be shown that this modeling of OCI includes that of \cite{J:Lozano_Limits, J:Heath_Distributed_Antenna_OCI_2011}. In addition, the $M$-dimension vector $\mathbf{n}_u$ represents the noise modeled as additive white Gaussian such that $\mathbb{E}\{\|\sum_{u\in\mathcal{U}_b} \mathbf{H}_{u,b}\mathbf{x}_{u,b}\|^2\}/\mathbb{E}\{\|\mathbf{n}_u\|^2\} = \snr$. We further normalize the channel matrices in order to have, in average, ICI power at the signal level as $\mathbb{E}\{\|\sum_{n\in\mathcal{U}_\ell}\mathbf{H}_{u,\ell}\mathbf{x}_{n,\ell}\|^2\} = \alpha(B-1)^{-1}\mathbb{E}\{\sum_{u\in\mathcal{U}_b}\|\mathbf{H}_{u,b}\mathbf{x}_{u,b}\|^2\}$ and OCI power at the signal level as $\mathbb{E}\{\|\mathbf{g}_u\|^2\} = \beta\mathbb{E}\{\sum_{u\in\mathcal{U}_b}\|\mathbf{H}_{u,b}\mathbf{x}_{u,b}\|^2\}$, where we have assumed that $\mathbb{E}\{\|\mathbf{H}_{u,b}\mathbf{x}_{u,b}\|^2\} = \mathbb{E}\{\|\mathbf{H}_{u,b}\mathbf{x}_{k,b}\|^2\}$ for $k\in\mathcal{U}_b$ with $k\neq u$. The system model of \eqref{Eq:Received_Signal} is capable of describing a wide range of interference scenarios by varying the parameters $\alpha$ and $\beta$ as well as the distribution of $\mathbf{g}_u$, thereby capturing how interference coordination might perform for MTs in different network setups. An example illustration of this model is depicted in Fig$.$~\ref{Fig:General_IBC}. The three BSs in the center of the figure are assumed to form a coordination cluster. The MTs falling into the regions covered by these BSs are subject to relative interference $\alpha$ from intra-cluster BSs, and aggregate interference $\beta$ from each of the out-of-cluster BSs.

\section{Advanced Coordinated Beamforming Schemes}\label{sec:CB_Schemes}
In this section, the system model of Section~\ref{sec:Cellular_Model} is first employed to a simplistic cellular network in order to demonstrate the theoretical gains of JT CoMP and CB schemes over representative non-coordinated ones as well as to compare JT CoMP with CB. Then, we present performance comparisons among CB schemes requiring full CSI exchange among coordinating BSs as well as schemes that operate with limited coordination overhead. The compared schemes differ on the considered design objective and the level of taking network interference under consideration.   

\subsection{Theoretical Gains Through An Example}\label{sec:Example}
We consider a cluster of $B=2$ BSs as a part of a large cellular network, which aims at serving $2$ MTs in every time-frequency resource unit; one MT is associated to the one BS and the other MT to the other BS. Focusing on the presented system model and using the classical bounds for the individual MT rates in multiple-input single-output (MISO) IFCs \cite{Gesbert2010}, it holds that:   
\begin{figure}[!t]
\centering
\includegraphics[width=4.9in]{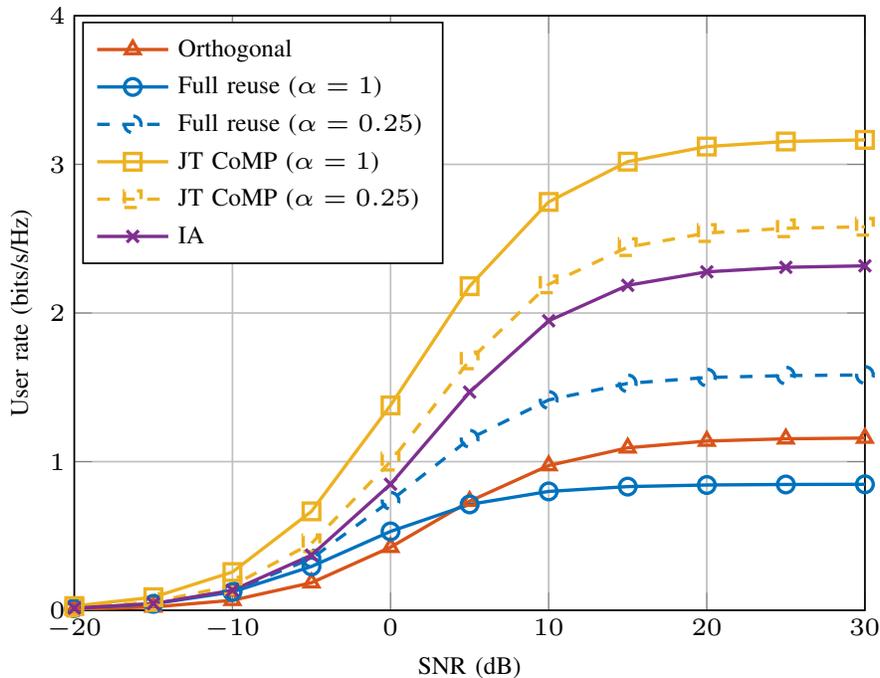}
\caption{Theoretical comparison of JT CoMP and IA as well as full reuse and orthogonal sharing of time-frequency resources for the downlink of a cluster of a cellular network comprised of $2$ multi-antenna BSs and $1$ single-antenna MT assigned per BS. The case where $\alpha = 1$ is suitable for describing cell-edge MTs which are subject to ICI having the same relative strength with their intended signal. In both cases, $\beta=0.25$ with respect to the power of the intended signal.}
\label{Fig:Comparisons_Toy}
\end{figure} 
\begin{itemize}
	\item With full reuse of time-frequency resources, each MT is subject to interference from every BS not associated with, and its rate is upper bounded as $\log_2(1+\snr/(\alpha\snr+\beta\snr+1))$;
	\item With orthogonal allocation of the resources, ICI is absent but the prelog factor $0.5$ appears on each individual MT rate, yielding $0.5\log_2(1+\snr/(\beta\snr +1))$;
	\item With the CB scheme based on interference alignment (IA) \cite{J:Iterative_Jafar}, ICI can be completely nulled, and the individual MT rate becomes $\log_2(1+\snr/(\beta\snr+1))$; and
	\item With ideal JT CoMP, the interference power actually boosts the intended signals and the individual MT rate is given by $\log_2(1+(1+\alpha)\snr/(\beta\snr+1))$. 	
\end{itemize} 
The latter rates for each individual MT are sketched in Fig$.$~\ref{Fig:Comparisons_Toy} with OCI being $6$ dB lower than that of the power of the intended signal, i$.$e$.$, $\beta=0.25$, and for two different values of $\alpha$, which reveals the relative power of ICI. As expected, both coordinated transmission schemes provide substantial gains compared with full reuse and orthogonal transmission when the network operates in the interference-limited regime, i$.$e$.$, when $\snr$ increases. As $\alpha$ approaches $0 $ the gain of JT CoMP over IA decreases. For example, for $\snr=15$ dB and $\alpha=1$ in Fig$.$~\ref{Fig:Comparisons_Toy}, IA results in a nearly $100\%$ gain over orthogonal transmission while, this gain becomes nearly $180\%$ for JT CoMP. When $\alpha$ decreases, the latter gain of IA remains the same whereas, that of JT CoMP decreases to nearly $110\%$. This example illustrates that, in many cases of interest, a large part of the coordination gain comes more from the removal of interference from the signal of interest rather than from stacking the powers of multiple BSs. It is also noted that, when considering practical implementation issues in achieving JT CoMP, the bonus of full coordination becomes even lower, since JT CoMP is more afflicted by degraded CSI and dirty RF than CB \cite{Jungnickel2014}.

\subsection{CB Schemes with Full CSI Exchange}\label{sec:Full_Coordination}
\begin{figure}[!t]
\centering
\includegraphics[width=4.9in]{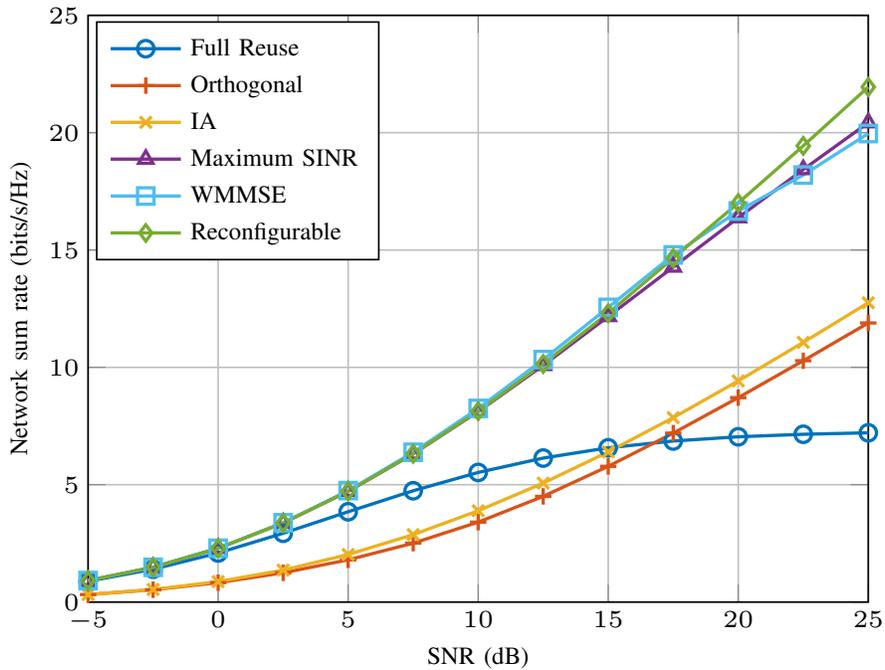}
\caption{Achievable sum rates for different CB schemes with full CSI exchange and $\alpha=1$ and $\beta=0$. The coordination cluster comprises of $3$ $4$-antenna BSs and $1$ $2$-antenna MT associated with each BS. A maximum of $10$ iterations was used for each of the iterative schemes maximum SINR, WMMSE, and Reconfigurable. The performance of full reuse and orthogonal $4\times 2$ MIMO transmission is also depicted.}
\label{Fig:IFC_Results_1}
\end{figure}
\begin{figure}[!t]
\centering
\includegraphics[width=4.9in]{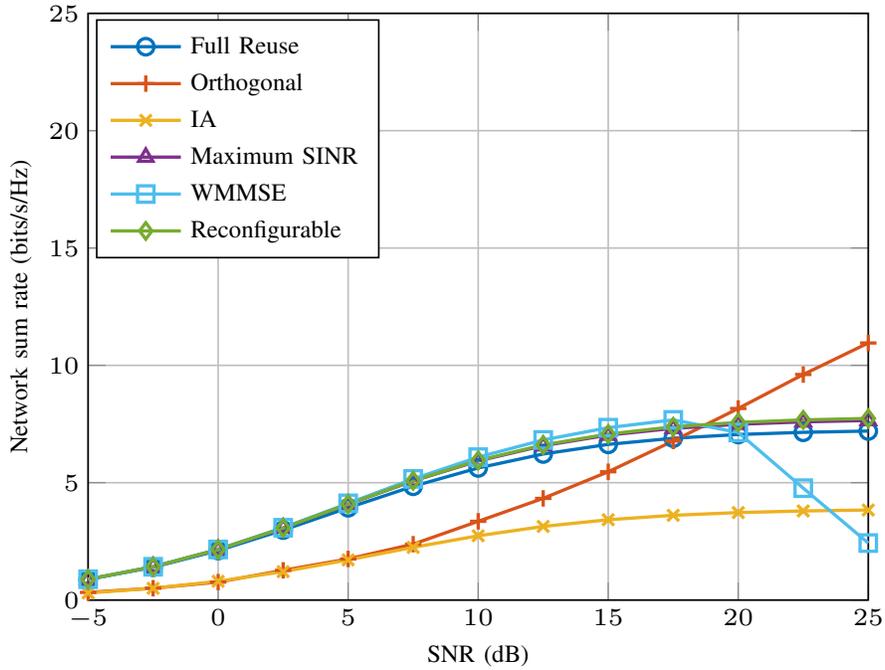}
\caption{Achievable sum rates for different CB schemes with full CSI exchange and $\alpha=0.25$ and $\beta=0.25$. The coordination cluster comprises of $3$ $4$-antenna BSs and $1$ $2$-antenna MT associated with each BS. A maximum of $10$ iterations was used for each of the iterative schemes maximum SINR, WMMSE, and Reconfigurable. The performance of full reuse and orthogonal $4\times 2$ MIMO transmission is also depicted.}
\label{Fig:IFC_Results_2}
\end{figure}

We hereinafter focus on the $B$-user $N\times M$ IFC, which constitutes a special case of the system model of Section~\ref{sec:Cellular_Model} where each $\mathcal{U}_b$ comprises of exactly one MT. In Figs$.$~\ref{Fig:IFC_Results_1} and~\ref{Fig:IFC_Results_2}, we consider a coordination cluster of $B=3$ BSs with $N=4$ and $M=2$, and compare the ergodic performance with optimum receivers for different values of $\alpha$ and $\beta$, and spatially independent Rayleigh fading of the following CB schemes: \textit{i}) IA \cite{J:Iterative_Jafar} that aims at aligning, and then nulling interference at each MT belonging in the BS cluster; \textit{ii}) Maximum signal-to-interference-plus-noise ratio (SINR) \cite{J:Iterative_Jafar} that targets at maximizing the received SINR of each transmitted information data stream in the cluster; \textit{iii}) Weighted Minimum Mean Squared Error (WMMSE) \cite{J:Luo_TSP_IterMMSE} that minimizes a metric for the whole network that is based on the MMSE; and \textit{iv}) Reconfigurable \cite{J:Alexandg_Recon2013}. The latter scheme combines a network-wide MMSE criterion with the single-user MIMO waterfilling solution in order to maximize the rate of each MT associated with the coordination cluster, accordingly to the condition of its desired channel and the whole network's interference level. Although, for all aforementioned CB schemes, we consider here a centralized implementation with full CSI exchange among coordinating BSs, it is noted that, for the maximum SINR, WMMSE, and the Reconfigurable schemes, distributed versions are also available, where explicit CSI exchange among BSs is avoided, and thus, coordination overhead can be potentially reduced.

The IA, maximum SINR, WMMSE, and the Reconfigurable CB schemes are linear schemes, which means that each BS transmits its signal using precoded symbols as $\mathbf{x}_{u,b}=\mathbf{V}_b\mathbf{s}_b$, where $\mathbf{V}_b$ represents the $N\times d_b$ precoding matrix and $\mathbf{s}_b$ is the $d_b$-dimension information stream vector. Upon signal reception, each MT estimates the desired transmitted symbols using a $d_b \times M$ decoding matrix $\mathbf{U}_b$, forming $\mathbf{\hat s}_b = \mathbf U_b \mathbf y_b$. For IA and the maximum SINR schemes in Figs$.$~\ref{Fig:IFC_Results_1} and~\ref{Fig:IFC_Results_2}, each $d_b$ was set to $0.5\min(N,M)=1$ according to the IA feasibility conditions \cite{J:Iterative_Jafar}, and $\mathbf{V}_b$ for all $b$ was obtained in closed form for IA and iteratively for maximum SINR. For both the iterative schemes WMMSE and Reconfigurable, each $d_b$ was initialized as $d_b=\min(N,M)=2$ and obtained at the end of the algorithmic iterations or upon convergence, explicitly for the Reconfigurable scheme and implicitly for WMMSE together with all $\mathbf{V}_b$ matrices. More specifically, the Reconfigurable scheme outputs $d_b$ to be sent by each coordinated BS $b$ together with their beamforming directions whereas, WMMSE only generates the transmit covariances matrices with possibly some streams set to zero power, and thus unusable. This means that, for the latter scheme the optimum $d_b$ needs to be searched in some way, a fact that will cause an extra overhead in practical networks and possibly decrease performance. As it can be concluded from Figs$.$~\ref{Fig:IFC_Results_1} and~\ref{Fig:IFC_Results_2} for a maximum of $10$ iterations per iterative scheme, the performance of all considered CB schemes is susceptible to ICI and OCI. This behavior for OCI was also observed in \cite{Mungara2015} for IA. For example, for $\snr=15$ dB, it is shown that the performance of all CB schemes drops approximately $45\%$ between the two interference scenarios, according to which $\alpha$ decreases from $1$ to $0.25$ and $\beta$ increases from $0$ to $0.25$. Interestingly, for the considered interference cases in both figures and $\snr<17.5$ dB, the maximum SINR, WMMSE, and Reconfigurable schemes, that take OCI interference under consideration, provide equal to or slightly more than $100\%$ improvement compared to IA. This behavior witnesses that maximum SINR, WMMSE, and Reconfigurable schemes are highly resilient to the $\alpha$ values, however, their resilience to $\beta$ values is low, especially for WMMSE and high $\snr$ values. This result tends to reinforce the necessity of considering OCI when designing CB schemes, and justify their study under practical network conditions. As also demonstrated in Fig$.$~\ref{Fig:IFC_Results_2}, the majority of the CB schemes perform very close or slightly better to full reuse and for $\snr<17.5$ dB, all CB schemes outperform orthogonal MIMO transmissions. However, for $\snr>17.5$ dB, orthogonal transmissions is the best option, a fact that witnesses that to achieve the best performance for general values of $\beta$, the coordination cluster needs to adopt a dual-mode operation, which switches depending on the $\beta$ values between the Reconfigurable CB scheme for example and orthogonal MIMO transmissions. 

In the CB schemes discussed before, MTs served by the clustered BSs are assumed to be clustered so as to create a separate group. This transpires in the current LTE standard, in particular, LTE release $11$ describes a \emph{CoMP cluster} in which BSs may coordinate their transmissions \cite{3GPPComPV11}. This CoMP cluster forms the basis into which the techniques \cite{J:Iterative_Jafar, J:Luo_TSP_IterMMSE, Suh2011, J:Alexandg_Recon2013} may be implemented, although as we will discuss in the following section, information exchange between the coordinated BSs is still not standardized. Inside a CoMP cluster, a MT may estimate the channels of its interferers through specific CSI structures and commands. This CSI may then be used to compute interference-aware receive filters \cite{3GPPIRCspecs} or fed back to their associated BS for further processing. The BSs inside a CoMP cluster may also be able to exchange CSI when operating in time-division duplexing mode \cite{Irmer2011}, by making use of the channel reciprocity property that is lacking in the more common frequency-division duplexing (FDD) mode. Notwithstanding, CSI exchange among coordinated BSs is still a complex operation; it weighs heavily on the backbone network \cite{Jungnickel2014}, and as of today, there is also no specific standardized mechanism on how or when to transfer this information. Therefore, the straightforward implementation of the described CB schemes outside of a vendor-locked configuration is still out of reach.

\subsection{CB Schemes with Limited Coordination Overhead}\label{sec:Low_Overhead}
One example of a CB scheme with limited coordination overhead is the Downlink IA presented in \cite{Suh2011}, which is suitable for the more general IBC. This schemes capitalizes on the standards' specifications \cite{3GPPComPV11} to allow each MT to estimate its strongest interferer, and feedback good precoder candidates to its associated BS. Consider a cellular network with coordination clusters of $B=2$ BSs, where each coordinated BS aims at sending a single information stream to its $K=N-1$ associated MTs. The basic idea of Downlink IA is to force the received signal at each MT associated to the cluster from the non-intended coordinated BS in a signal subspace of rank $N-1$, thus freeing up one decoding dimension from interference for the desired signal. These decoding directions create equivalent MISO channels for the MTs associated to the cluster, which can then feedback them to their associated BS. Each BS can then employ any multi-user MIMO technique \cite{Gesbert2010} to multiplex the $N-1$ information streams towards its respective MTs, such as zero-forcing beamforming. The benefit of the Downlink IA scheme is that, although each BS only frees a single dimension, the interference-free direction is different for each MT, thereby enabling multi-user diversity. One can contrast Downlink IA with FFR schemes \cite{Hamza2013}, where the dimension freed in frequency is the same for all MTs. In our performance evaluation, we further assume that each MT can learn the precoder chosen by the BS for their stream at the end. Since they have already estimated the channel from the non-intended clustered BS, they know the necessary information to update their receivers to Interference Rejection Combining (IRC) receivers, as described in \cite{3GPPIRCspecs}.
\begin{figure}[!t]
\centering
\includegraphics[width=4.9in]{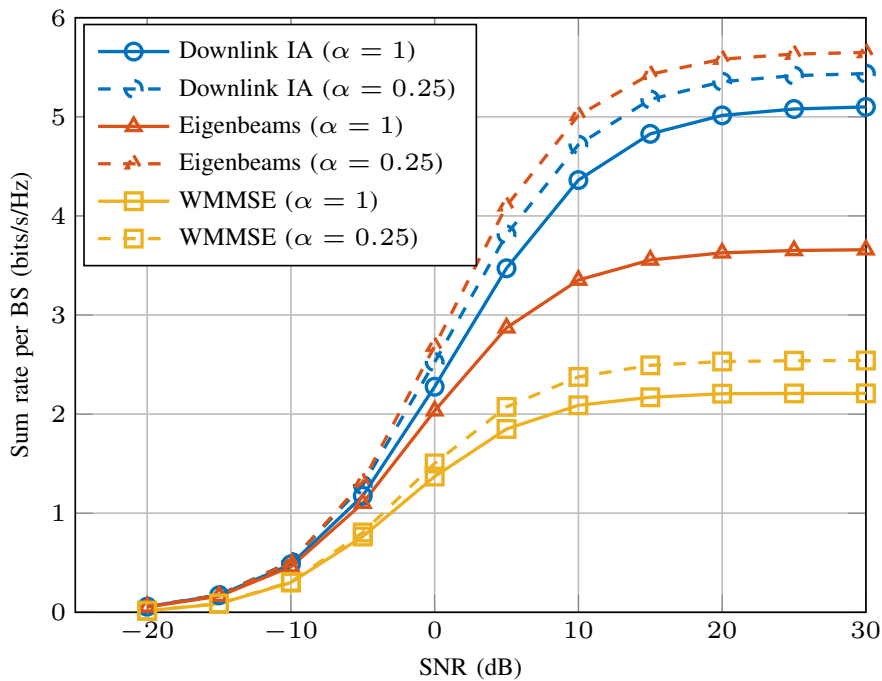}
\caption{Achievable rates per BS for different values of $\alpha$ and $\beta=0.25$ for a IBC with $4$-antenna nodes and a main interferer for each coordinated BS. For both the Eigenbeams and Downlink IA schemes, IRC receivers have been used. The performance of the WMMSE scheme is also illustrated for comparison purposes.}
\label{Fig:IBC_Results}
\end{figure}

The performance of Downlink IA with IRC receivers over spatially independent Rayleigh fading channels is illustrated in Fig$.$~\ref{Fig:IBC_Results} as a function of the $\snr$ for different values of $\alpha$, $\beta=0.25$, $B=2$, $M=N=4$, and $K=3$. Within this figure, we also plot the performance of a more classical multi-user MIMO scheme where interference is not exploited, and for which the decoding direction of each MT assigned to the cluster is chosen as the strongest eigenvector of its intended channel; this scheme is denoted as the Eigenbeams scheme. Note that with the Eigenbeams scheme, each BS can support $4$ MTs whereas, with Downlink IA each BS served $3$ MTs. As seen from the figure, and as expected, for cell-edge MTs there is potentially a very large SINR gain coming from the removal of the interfering coordinated BS. In that case, Downlink IA shows a $50\%$ gain from the Eigenbeams scheme in the average sum rate per coordinated BS. On the other hand, the gain of Downlink IA for MTs that are not at the cell edge, and thus do not experience a strong interferer, is reduced. As highlighted by the theoretical example in Section~\ref{sec:Example}, the performance of Downlink IA depends to the values of both $\alpha$ and $\beta$; if $\alpha$ is much larger than $\beta$, we have a strong interest in removing the interference even if it means being somewhat misaligned with our own channel. On the other hand, if the remaining OCI is on the level of the ICI, Downlink IA provides less gains than a straightforward multi-user MIMO scheme like Eigenbeams. This is in line with recent analyses, as e$.$g$.$ in \cite{Mungara2015}, where it was shown that blindly applying IA in a clustered cellular network is altogether detrimental. We can also conclude from Fig$.$~\ref{Fig:IBC_Results} that the performance of the WMMSE scheme is poor in this context, since it targets at minimizing the interference from the non-intended clustered BS even if it has to shut down transmissions to its MTs. At convergence of this iterative algorithm, a subset of the MTs will experience a very high SINR, but since some streams will be unused, the overall performance is lower than that of Downlink IA or Eigenbeams.

\section{Coordinated Transmission with the LTE Feedback Specification}\label{sec:Standard_Feedback}
The CB schemes presented in the previous section necessitate sort of CSI exchange among the BSs belonging in the coordination cluster. However, there is still no standardized mechanism in the current LTE specifications for full CSI exchange in cellular networks. This means that non-proprietary attempts at achieving CB are not truly possible as of today. As such, CB is not feasible outside of a vendor-locked coordinated set of BSs, or within a single BS with remote radio heads. This precludes many of the presented advanced CB schemes, which require CSI exchange and possibly joint computation of the transmission parameters among the coordinated BSs.
\begin{figure}[!t]
\centering
\includegraphics[width=4.9in]{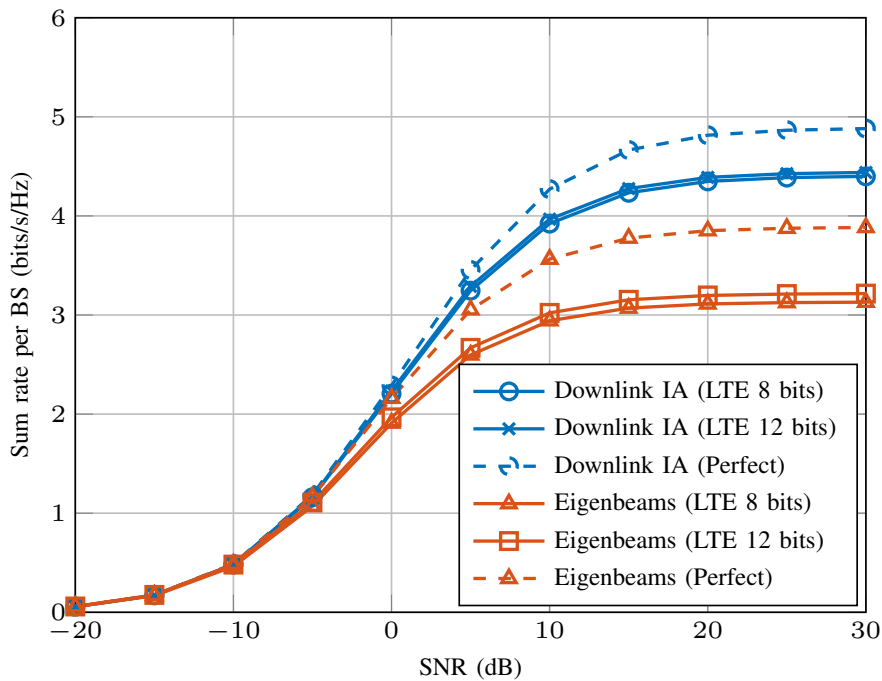}
\caption{Impact of non-ideal standardized feedback on the achievable sum-rate performance per coordinated BS with the Downlink IA and Eigenbeams schemes for a IBC with $4$-antenna nodes and a main interferer for each coordinated BS. In all cases, $\alpha=1$ and $\beta=0.25$, and MTs are assumed to compute the optimal IRC receivers having knowledge of the precoder chosen by their assigned BS.}
\label{Fig:Standard_IBC}
\end{figure}

Focusing on the LTE release $12$ for feedback specifications, we henceforth compare the performance of the Downlink IA and Eigenbeams schemes for the IBC scenario of Fig$.$~\ref{Fig:IBC_Results} under standardized feedback, and compare it with the ideal feedback case. In particular, the CSI feedback needed in these schemes is limited to the feedback of only a channel quality indicator (CQI) and a precoding matrix indicator (PMI) for each frequency subband. The physical layer procedures related to this feedback and the PMI codebooks are described in \cite{3GPPPhysProc}. In Fig$.$~\ref{Fig:Standard_IBC}, we evaluate the performance of Downlink IA and Eigenbeams with practical feedback using $4$-antenna network nodes and a $8$-bit codebook that creates a family of $256$ possible precoders. To apply this codebook to the considered IBC scenario, we feedback the equivalent channel by using the PMI to the closest precoder in the family. As depicted from Fig$.$~\ref{Fig:Standard_IBC}, this procedure results into a net performance loss of about $10\%$ for the Downlink IA scheme and $20\%$ for the Eigenbeams scheme. It can be shown that this loss is not entirely linked to the somewhat coarse feedback quantization, but rather in the way the codebook is constructed in \cite{3GPPPhysProc}. In fact, increasing the number of bits in the feedback scheme, while keeping the same codebook construction, does not improve the performance substantially. This indicates that the sheer number of bits for the feedback channel is not itself the strongest indicator of feedback quality, and that codebook construction is in fact a fundamental question. Higher precision in the feedback process as well as accurate CSI estimation are thus still two of the key questions to answer today for coordinated transmissions schemes as well as for many other channel-dependent signal processing techniques. In addition, practical CSI exchange between BSs participating in a coordination cluster is undefined as of the latest LTE release. There is actually no standardized way of encoding CSI in FDD systems. The specifications of the backbone communications in a CoMP set are also left to vendor implementations, precluding any inter-vendor CoMP set to be set up in practice.

\section{Conclusion and Research Directions}\label{sec:Conclusions}
As network deployments become denser, interference arises as a dominant performance degradation factor that is almost irrespective to the underlining physical-layer technology. The feature of coordinating BS transmissions to manage interference in cellular networks is already a part of the latest LTE release, offering significant potential for performance improvement especially at the cell edges. Among the recently proposed coordination schemes, there exist CB schemes that require coordination overhead that is more or less compatible with the current standard's specifications, and adapt satisfactory to ICI, while showing some resilience to OCI. However, to maximize the benefit from CB in future communication networks, certain advances need to take place. One of these is BS clustering that needs to be both dynamic and scalable. Efficient clustering methods, based for example on network connectivity or received SINR, that keep OCI levels to the minimum can be combined with CB schemes to boost network performance. Another necessary progress in coordination schemes is the design of techniques for information exchange with low overhead among the coordinated network nodes. The coordination overhead of the latest CB schemes is still far from what can be supported in the current LTE release. This necessity becomes even more prominent in fully distributed CB schemes, where information needs to be exchanged iteratively between transmitters and receivers. In fact, it is yet unclear how to practically implement iterative CB schemes and their required information exchange overhead. There are issues in both the actual form the information messages will take, the structure of the message-passing shells, and most importantly the quantization that has to be done on the message content. Up to this day, there is little research on designing CB schemes where the iterative computation supports noisy or quantized messages. This is also related to the accuracy of CSI as measured by members of the coordination cluster. Last but not least, coordination schemes need to be designed to account for the characteristics of technologies intended for next generation networks, such as for example full duplex radios and massive MIMO with possibly hybrid analog and digital processing. 

\bibliographystyle{IEEEtran}

\end{document}